\def\half{\frac{1}{2}}

\def\mb#1{\mbox{\boldmath{$#1$}}}

\documentclass[amsmath,amssymb,showkeys, showpacs]{revtex4}
\usepackage{graphicx}
\begin{document}

\hspace*{4 in}CUQM-141\\
\vspace*{0.4 in}
\title{Discrete spectra for confined and unconfined $-a/r + br^2$ potentials in $d$-dimensions}

\author{Richard L. Hall$^1$, Nasser Saad$^2$, and K. D. Sen$^3$}
\address{$^1$ Department of Mathematics and Statistics, Concordia University,
1455 de Maisonneuve Boulevard West, Montr\'eal,
Qu\'ebec, Canada H3G 1M8}\email{rhall@mathstat.concordia.ca}
\address{ $^2$ Department of Mathematics and Statistics,
University of Prince Edward Island, 550 University Avenue,
Charlottetown, PEI, Canada C1A 4P3.}\email{nsaad@upei.ca}
\address{$^3$ School of Chemistry, University  of Hyderabad 500046, India.}
\email{sensc@uohyd.ernet.in}

\begin{abstract}
\noindent Exact solutions to the $d$-dimensional Schr\"odinger equation, $d\geq 2$, for Coulomb plus harmonic oscillator potentials $V(r)=-a/r+br^2$, $b>0$ and $a\ne 0$ are obtained. The potential $V(r)$ is considered both in all space, and under the condition of spherical confinement inside an impenetrable spherical box of radius $R$.  With the aid of the asymptotic iteration method, the exact analytic solutions under certain constraints, and general approximate solutions, are obtained. These exhibit the parametric dependence of the eigenenergies on $a$, $b$, and $R$. The wave functions have
the simple form of a product of a power function, an exponential function, and a polynomial. In order to achieve our results the question of determining the polynomial solutions of the second-order differential equation
$$\left(\sum_{i=0}^{k+2}a_{k+2,i}r^{k+2-i}\right)y''+\left(\sum_{i=0}^{k+1}a_{k+1,i}r^{k+1-i}\right)y'-\left(\sum_{i=0}^{k}\tau_{k,i}r^{k-i}\right)y=0$$
for $k=0,1,2$ is solved. 

\end{abstract}

\maketitle
\section{Introduction}\label{intro}
\subsection{Formulation of the problem in $d$ dimensions}
\noindent The $d$-dimensional Schr\"odinger equation, in atomic units $\hbar=\mu=1$, with
a spherically symmetric potential $V(r)$ can be written as
\begin{equation}\label{eq1}
\left[-{1\over 2}\Delta_d +V(r)\right]\psi(r)=E\psi(r),
\end{equation}
where $\Delta_d$ is the $d$-dimensional Laplacian operator and $r^2=\sum_{i=1}^d x_i^2$.
Following \cite{louck}, in order to transform (\ref{eq1}) to the $d$-dimensional spherical coordinates $(r, \theta_1,\theta_2,\dots,\theta_{d-1})$, we separate variables using
\begin{equation}\label{eq2}
\psi(r)=r^{(d-1)/2}u(r)Y_{l_{d-1}\dots l_1}(\theta_1\dots\theta_{d-1}),
\end{equation}
where $Y_{l_{d-1}\dots l_1}(\theta_1\dots\theta_{d-1})$ is a normalized spherical harmonic with characteristic value $l(l+d-2), l=0,1,2,\dots$ (the angular quantum numbers), one obtains the radial Schr\"odinger equation as
\begin{equation}\label{eq3}
\left[-{1\over 2}\left({d^2\over dr^2}-{(k-1)(k-3)\over 4r^2}\right)+V(r)-E\right]u(r)=0,\quad\quad \int_0^\infty u^2(r)dr=1, u(0)=0,
\end{equation}
where $k=d+2l$. Assume that the potential $V(r)$ is less singular than the centrifugal term so that
$$u(r)\sim r^{1\over 2}(k-1)\quad\quad (r\rightarrow 0).$$
We note that the Hamiltonian and boundary conditions of (\ref{eq3}) are invariant under the transformation 
$$(d,l)\rightarrow (d\mp2,l\pm 1).$$ Thus, given any solution for fixed $d$ and $l$, we can immediately generate others for different values of $d$ and $l$. Further, the energy is unchanged if $k=2\ell+d$ and the number of nodes $n$ is constant. Repeated application of this transformation produces a large collection of states, the only apparent limitation being a lack of interest in some values of $d$ (see, for example \cite{Doren}). In the present work, we consider the Coulomb plus a
harmonic oscillator potential
\begin{equation}\label{eq4}
V(r)=-{a\over r}+br^2,\quad\quad b>0
\end{equation} 
where $r=\|\mb{r}\|$ denotes the hyper-radius, and the coefficients $a$ and $b$ are both constant.
\medskip
\subsection{Degeneracy in spherically confined $d$-dimensional quantum model systems}
Since the early days of quantum mechanics there has been interest
in studying the Schr\"{o}dinger equation with model systems in higher
spatial dimensions \cite{Fock,alliluev58,Avery,Avery2}. The
so-called accidental degeneracy of the hydrogen atom and isotropic
harmonic oscillator, characterized by different sets of parity
conditions, is generally understood in terms of the corresponding
$SO(4)$ and $SU(3)$ symmetry groups \cite{Singer, Fradkin}. Following the introduction of
`interdimensional degeneracies' \cite{Herrick,HS} there have been
several reports involving arbitrary $d$-dimensional analyses 
covering many branches of chemical physics which have been briefly
reviewed in Refs.~\cite{Chatt,Dunn,Ma,Night}. It is interesting to note
here that the information-theoretical uncertainty-like
relationships in terms of the Shannon entropy \cite{Shannon,BBM}
and the Fisher measure \cite{Fisher,Dehesa} are also stated in
 $d$-dimensional form.
\vskip0.1true in
Owing to the recent interest in quantum dots and fullerine
encapsulated electronic systems there has been an upsurge of
interest in studying model quantum systems confined inside an
impenetrable sphere of radius $R$. We shall present here a brief
description of the new degeneracy-related changes which are known
to occur in the $d$-dimensional H atom $V_c=-a/r$ and the isotropic harmonic
oscillator $V_h=br^2$. The eigenspectrum of the spherically confined H atom
(SCHA) is characterized by three kinds of degeneracy 
\cite{pupyshev98}. Two of them are generated from the specific
choice of the radius of confinement $R$, chosen exactly at the
radial nodes corresponding to the free hydrogen atom (FHA) wave
functions. In the \emph{incidental degeneracy} case, the confined
$(\nu,\ell)$ state with the principal quantum number $\nu$ is
iso-energic with $(\nu+1,\ell)$ state of the FHA with energy
$-1/\{2(\nu+1)^2\}$ atomic units (a.u.), at an $R$ defined by the radial node in the
FHA. For example, the $(\nu,\ell)$ state corresponding to the
lowest energy value, when confined at the radius $R$ given by the
radial node the first excited \emph{free} state $(\nu+1,\ell)$,
increases in such a way that the confined-state energy becomes the same as
excited free-state energy. The specific node in question is given
by $R=0.24({2\ell}+d-1)({2\ell}+d+1)$. Such a degeneracy can be
realized at similar choices for $R$ where multiple nodes exist in
the second and higher excited states of a given $\ell$. However,
such closed analytical expressions for the radial nodes are not
available in the case of higher excited states. In the
\emph{simultaneous-degeneracy} case, on the other hand, for all
$\nu \ge \ell+2$, each  pair of confined states denoted by
$(\nu,\ell)$ and $(n+1,\ell+2)$ state, confined at the common
$R=0.24({2\ell}+d-1)({2\ell}+d+1)$, become degenerate. Note that
the pair of levels in the free state are nondegenerate. Both these
degeneracies have been shown \cite{pupyshev98} to result from the
Gauss relationship applied at a unique $R_c$ by the confluent
hypergeometric functions that describe the general solutions of
the SCHA problem. Finally, the interdimensional
degeneracy \cite{Herrick,HS} arises, as in the case of the free
hydrogen atom, due to the invariance of the Schr\"{o}dinger
equation to the transformation
$(\ell,d)~\rightarrow~(\ell\pm~1,~d\mp~2)$. In order to preserve
the number of nodes in the radial function, it is simultaneously
necessary to make the transformation $\nu~\rightarrow~ \nu + 1$.
The \emph{incidental degeneracy} observed in the case of a
spherically confined isotropic harmonic oscillator (SCIHO) is
qualitatively similar to that of the SCHA. For example, the only
radial node in the first excited free state of any given $\ell$
for $d$-dimensional SCIHO is located at $R=\sqrt{(2\ell+d)/2}$. For
the multiple node states, the corresponding numerical values must
be used. However, the behavior of the two confined states at a
common radius of confinement is found to be interestingly
different \cite{sen,spm}. In particular, for the SCIHO the pairs
of the confined states defined by $(\nu=\ell+1,\ell)$ and
$(\nu=\ell+2,\ell+2)$ at the common $R=\sqrt{(2\ell+d)/2}$ a.u.,
display for all $\nu$, a constant energy separation of
\emph{exactly} $2$ harmonic-oscillator units,~$2\hbar \omega$ ,
with the state of higher $\ell $ corresponding to the lower energy. It
is interesting to note that the two confined states at the
common $R$ with $\Delta \ell=2$, considered above
 contain different numbers of radial nodes.
 The condition for interdimensional
degeneracy\cite{Herrick,HS} due to the invariance of the
Schr\"{o}dinger equation remains the same as before. Recently, the
confined systems of the $d$-dimensional hydrogen atom \cite{Jaber}
and harmonic oscillator \cite{Ed} have been studied.
Problems involving short-range potentials in $d$ dimensions have recently been considered \cite{gusun,agboola}. 
 In the light of the above discussion, it is interesting to study the various
aforementioned degeneracies in the free and spherically confined
$d$-dimensional potential generally given by $V(r) = V_c+V_h=-{a}/{r} +
b r^2$.
\medskip
\subsection{Organization of the paper}
\noindent The present paper is organized as follows. In section 2, we discuss some general spectral features and bounds, in section~3 we briefly review the asymptotic iteration method of solving a second-order linear differential equation where we discuss the necessary and sufficient conditions for certain classes of differential equations with polynomial coefficients to have polynomial solutions. In sections 4 and 5, we use the asymptotic iteration method (AIM) to study how the
eigenvalues depend on the potential parameters $a, b, R$, repectively for the free system ($R = \infty$), and for finite $R$.
In each of these sections, the results obtained are of two types: exact analytic results that are valid when certain
parametric constraints are satisfied, and accurate numerical values for arbitrary sets of potential parameters. 
\section{Some general spectral features and analytical energy bounds}\label{bounds}
We shall show shortly that the Hamiltonian $H$ is bounded below. The eigenvalues of $H$ may therefore be characterized variationally.  The eigenvalues $E_{n,\ell}^d = E(a,b,R)$ are monotonic in each parameter. For $a$ and $b$, this is a direct consequence of the monotonicity of the potential $V$ in these parameters. Indeed, since $\partial V/\partial a = -1/r <0$ and $\partial V/\partial b = r^2 > 0$, it follows that
\begin{equation}\label{monotoneEab}
\frac{\partial E(a,b,R)}{\partial a} < 0\quad {\rm and}\quad\frac{\partial E(a,b,R)}{\partial b} > 0.
\end{equation}
The monotonicity with respect to the  box size $R$ may be  proved by a variational argument.
 Let us consider two box sizes, $R_1 < R_2$ and an angular momentum subspace labelled by a fixed $\ell.$
We extend the domains of the wave functions in the finite-dimensional subspace spanned by the first $N$ radial eigenfunctions for $R = R_1$ so that the new space $W$ may be used to study the case $R = R_2$. We do this by defining the extended eigenfunctions so that $\psi_i(r) = 0$  for $R_1 \le r\le R_2.$  We now look at $H$ in $W$ with box size $R_2$.  The minima of the energy matrix $[(\psi_i,H\psi_j)]$ are the exact eigenvalues for $R_1$ and, by the Rayleigh-Ritz principle, these values are one-by-one upper bounds to the eigenvalues for $R_2.$ Thus, by  formal argument we deduce what is perhaps intuitively clear, that the eigenvalues increase as $R$ is decreased, that is to say
\begin{equation}\label{montoneER}
\frac{\partial E(a,b,R)}{\partial R} < 0.
\end{equation}
From a classical point of view, this Heisenberg-uncertainty effect is perhaps counter intuitive: if we try to squeeze the electron into the Coulomb well by reducing $R$, the reverse happens; eventually, the eigenvalues become positive and arbitrarily large, and less and less affected by the presence of the Coulomb singularity.
\vskip0.1true in
\noindent For some of our results we shall consider the system unconstrained by a spherical box, that is to say $R = \infty.$ For these cases, we shall write $E_{n\ell}^d = E(a,b).$  If a very special box is now considered, whose size $R$ coincides with any radial node of the $R=\infty$ problem, then the two problems share an eigenvalue exactly. This is an example of a very general relation which exists between constrained and unconstrained eigensystems, and, indeed, also between two constrained systems with different box sizes.

\medskip
The generalized Heisenberg uncertainty relation  may be expressed  \cite{GS,RS2} for dimension $d\ge 3$  as the operator inequality $-\Delta > (d-2)^2/(4r^2).$ This allows us to construct the following lower energy bound
\begin{equation}\label{lbound}
E > {\mathcal E} = \min_{0 <r \le R}\left[\frac{(d-2)^2}{8r^2} - \frac{a}{r} + b r^2\right].
\end{equation}
Provided $b \ge 0,$ this lower bound is finite for all $a$.  It also obeys the same scaling and  monotonicity laws as $E$ itself. But the bound is weak. For potentials such as $V(r)$ that satisfy $\frac{d}{dr}(r^2\frac{dV}{dr}) > 0,$  Common has shown \cite{common} for the ground state in $d=3$ dimensions that $\langle-\Delta\rangle > \langle 1/(2r^2)\rangle,$ but the resulting energy lower bound  is still weak.
\vskip0.1true in

For the unconstrained case $R = \infty$, however, envelope methods
\cite{env1,env2,env3,env4,env5,envcp} allow one to construct analytical upper and lower energy bounds with general forms similar to (\ref{lbound}). In this case we shall write $E_{n\ell}^d = E(a,b).$  Upper and lower bounds on the eigenvalues are based on the geometrical fact that $V(r)$ is at once a concave function $V(r) = g^{(1)}(r^2)$ of $r^2$ and a convex function $V(r) = g^{(2)}(-1/r)$ of $-1/r$.  Thus tangents to the $g$ functions are either shifted scaled oscillators above $V(r)$, or shifted scaled atoms below $V(r)$. The resulting energy-bound formulas are given by
\begin{equation}\label{ebounds}
\min_{r > 0}\left[\frac{1}{2 r^2} -\frac{a}{P_1r} + b (P_1r)^2\right]\, \le \,E_{n\ell}^d(a,b)\,\le\, \min_{r > 0}\left[\frac{1}{2 r^2} -\frac{a}{P_2r} + b (P_2r)^2\right],
\end{equation}
where (Ref. \cite{envcp2} Eqs.(1.11) and (1.12a))
\begin{equation}\label{P12}
P_1 = n+\ell+(d-1)/2 \quad {\rm and}\quad P_2 = 2n+\ell + d/2.
\end{equation}
We shall sometimes use also the convention of atomic physics in which, even for non-Coulombic central potentials, a principal quantum number $\nu$ is used and defined
by
\begin{equation}\label{nu}
\nu = n+ \ell + (d-1)/2,
\end{equation}
where $n = 0,1,2,\dots$ is the   number of nodes in the radial wave function.  It is clear that the lower energy bound has the Coulombic degeneracies, and the upper bound those of the harmonic oscillator. These bounds are very helpful as a guide when we seek very accurate numerical estimates for these eigenvalues.
\vskip0.1true in
\noindent Another related estimate is given by the `sum approximation' \cite{env5}
 which is more accurate than (\ref{ebounds}) and is known to be a lower energy bound for the bottom $E_{0\,\ell}^d$ of each angular-momentum subspace. The estimate is given by
\begin{equation}\label{sbound}
E_{n\ell}^d(a,b) \approx {\mathcal E}_{n\ell}^d(a,b) = \min_{r > 0}\left[\frac{1}{2 r^2} - \frac{a}{P_1r} + b (P_2r)^2\right].
\end{equation}
This energy formula has the attractive spectral interpolation property that it is {\it exact} whenever $a$ or $b$ is zero.
The energy bounds (\ref{ebounds}) and (\ref{sbound}) obey the same scaling and monotonicity laws is those of $E_{n\ell}^d (a,b).$  Because of their simplicity they allow one to extract analytical properties of the eigenvalues. For example, in Fig.~1 we show from Eq.(\ref{sbound}) approximately how the eigenvalue $E_{n\ell}^3(1,\half)$ depends  on $\ell$
for $n = 0,1,2.$

\begin{figure}
\centering
\includegraphics[height=6cm,width=9cm]{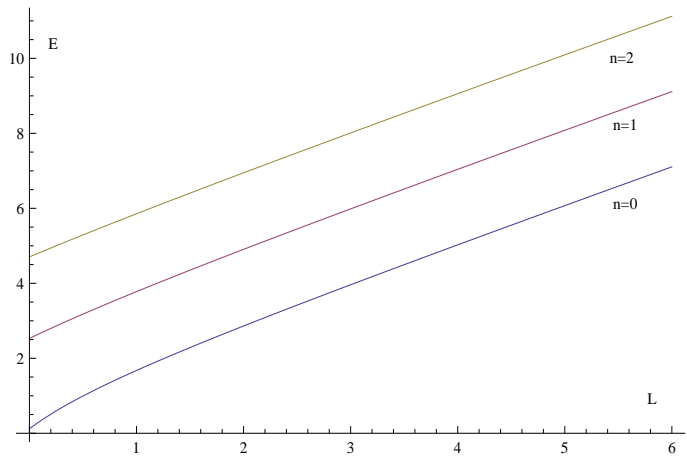}
\caption{The energy $E$ for $a=1$,~$b=\half$, $d=3$ as a function of $L= \ell$
for $n = 0,1,2.$}\label{fig1}
\end{figure}
  
\section{The asymptotic iteration method and some related results}\label{AIM}
\noindent The asymptotic iteration method (AIM) was originally
introduced \cite{aim} to investigate the solutions of differential
equations of the form
\begin{equation}\label{eq5}
y''=\lambda_0(r) y'+s_0(r) y,\quad\quad ({}^\prime={d\over dr})
\end{equation}
where $\lambda_0(r)$ and $s_0(r)$ are $C^{\infty}-$differentiable
functions. A key feature of this method is to note the invariant
structure of the right-hand side of (\ref{eq5}) under further
differentiation. Indeed, if we differentiate (\ref{eq5}) with
respect to $r$, we obtain
\begin{equation}\label{eq6}
y^{\prime\prime\prime}=\lambda_1 y^\prime+s_1 y
\end{equation}
where $\lambda_1= \lambda_0^\prime+s_0+\lambda_0^2$ and
$s_1=s_0^\prime+s_0\lambda_0.$ If we find the second derivative of
equation (\ref{eq5}), we obtain
\begin{equation}\label{eq7}
y^{(4)}=\lambda_2 y^\prime+s_2 y
\end{equation}
where $\lambda_2= \lambda_1^\prime+s_1+\lambda_0\lambda_1$ and
$s_2=s_1^\prime+s_0\lambda_1.$ Thus, for $(n+1)^{th}$ and
$(n+2)^{th}$ derivative of (\ref{eq5}), $n=1,2,\dots$, we have
\begin{equation}\label{eq8}
y^{(n+1)}=\lambda_{n-1}y^\prime+s_{n-1}y
\end{equation}
and
\begin{equation}\label{eq9}
y^{(n+2)}=\lambda_{n}y^\prime+s_{n}y
\end{equation}
respectively, where
\begin{equation}\label{eq10}
\lambda_{n}=
\lambda_{n-1}^\prime+s_{n-1}+\lambda_0\lambda_{n-1}\hbox{ ~~and~~
} s_{n}=s_{n-1}^\prime+s_0\lambda_{n-1}.
\end{equation}
From (\ref{eq8}) and (\ref{eq9}) we have
\begin{equation}\label{eq11}
\lambda_n y^{(n+1)}- \lambda_{n-1}y^{(n+2)} = \delta_ny {\rm
~~~where~~~}\delta_n=\lambda_n s_{n-1}-\lambda_{n-1}s_n.
\end{equation}
Clearly, from (\ref{eq11}) if $y$, the solution of
(\ref{eq5}), is a polynomial of degree $n$, then $\delta_n\equiv
0$. Further, if $\delta_n=0$, then $\delta_{n'}=0$ for all $n'\geq
n$. In an earlier paper \cite{aim} we proved the principal theorem
of AIM, namely 
\vskip0.1in

\noindent{\bf Theorem~1~\cite{aim}.} \emph{Given $\lambda_0$ and $s_0$ in
$C^{\infty}(a,b),$ the differential equation (\ref{eq5}) has the
general solution
\begin{equation}\label{eq12}
y(r)= \exp\left(-\int\limits^{r}{s_{n-1}(t)\over \lambda_{n-1}(t)} dt\right) \left[C_2
+C_1\int\limits^{r}\exp\left(\int\limits^{t}(\lambda_0(\tau) +
2{s_{n-1}\over \lambda_{n-1}}(\tau)) d\tau \right)dt\right]
\end{equation}
if for some $n>0$
\begin{equation}\label{eq13}
\delta_n=\lambda_n s_{n-1}-\lambda_{n-1}s_n=0.
\end{equation}
where $\lambda_n$ and $s_n$ are given by (\ref{eq10}).
}
\vskip0.1true in
\noindent Recently, it has been shown \cite{aim1} that the termination condition (\ref{eq13}) is necessary and sufficient for the differential
equation (\ref{eq5}) to have polynomial-type solutions of degrees at most $n$, as we may conclude from Eq.(\ref{eq11}). Thus, using Theorem 1, we can now find the necessary and sufficient conditions \cite{saad} for the polynomial solutions of the differential equation 
\begin{equation}\label{eq14}
(a_{3,0}r^3+a_{3,1}r^2+a_{3,2}r+a_{3,3})~y^{\prime \prime}+(a_{2,0}r^2+a_{2,1}r+a_{2,2})~y'-(\tau_{1,0} r+\tau_{1,1})~y=0,
\end{equation}
where $a_{k,j}, k=3,2,1, j=0,1,2,3$ are constants. These conditions are reported in the follow theorem.
\vskip0.1true in
\noindent{\bf Theorem 2~[\cite{saad} Theorem 5].} \emph{
The second-order linear differential equation (\ref{eq14}) has a polynomial solution of degree $n$ if 
\begin{equation}\label{eq15}
\tau_{1,0}=n(n-1)~a_{3,0}+n~a_{2,0},\quad n=0,1,2,\dots,
\end{equation}
provided $a_{3,0}^2+a_{2,0}^2\neq 0$ along with the vanishing of $(n+1)\times(n+1)$-determinant $\Delta_{n+1}$ given by
\begin{center}
$\Delta_{n+1}$~~=~~\begin{tabular}{|lllllll|}
 $\beta_0~~$ & $\alpha_1$ &$\eta_1$&~& ~&~ &~\\
  $\gamma_1$ & $\beta_1$ &  $\alpha_2$&$\eta_2$&~&~&~ \\
~ & $\gamma_2$  & $\beta_2$&$\alpha_3$&$\eta_3$&~&~\\
$~$&~&$\ddots$&$\ddots$&$\ddots$&$\ddots$&~\\ 
~&~&~&$\gamma_{n-2}$&$\beta_{n-2}$&$\alpha_{n-1}$&$\eta_{n-1}$\\
~&~&~&&~$\gamma_{n-1}$&$\beta_{n-1}$&$\alpha_n$\\
~&~&~&~&$~$&$\gamma_{n}$&$\beta_n$\\
\end{tabular}~~=~~0
\end{center}
where all the other entires are zeros and 
\begin{align}\label{eq16}
\beta_n&=\tau_{1,1}-n((n-1)a_{3,1}+a_{2,1})\notag\\
\alpha_n&=-n((n-1)a_{3,2}+a_{2,2})\notag\\
\gamma_n&=\tau_{1,0}-(n-1)((n-2)a_{3,0}+a_{2,0})\notag\\
\eta_n&=-n(n+1)a_{3,3}.
\end{align}
Here $\tau_{1,0}$ is fixed for a given $n$ in the determinant $\Delta_{n+1}=0$ (the degree of the polynomial solution). The coefficients of the polynomial solutions $y_n(r)=\sum_{i=0}^n c_i r^i$ satisfies the four-term recursive relation
\begin{align}\label{eq17}
&(i+2)(i+1)a_{3,3}c_{i+2}+\left[i(i+1)a_{3,2}+(i+1)a_{2,2}\right]c_{i+1}+\left[i(i-1)a_{3,1}+ia_{2,1}-\tau_{1,1}\right]c_i\notag\\
&+ \left[(i-1)(i-2)a_{3,0}+(i-1)a_{2,0}-\tau_{1,0}\right]c_{i-1}=0.
\end{align}
}
\vskip0.1true in
\noindent The results of this theorem go beyond the question of finding the polynomial solutions of the second-order linear differential equation
\begin{equation}\label{eq18}
(a_{2,0}r^2+a_{2,1}r+a_{2,2})~y^{\prime \prime}+(a_{1,0}r+a_{1,1})~y'-\tau_{0,0}~y=0.
\end{equation}
Indeed Eq.(\ref{eq18}) has a nontrivial polynomial solution of degree (exactly) $n\in \mathbb N$ (the set of nonnegative integers) if, for fixed $n$,
\begin{equation}\label{eq19}
\tau_{0,0}=n(n-1)~a_{2,0}+n~a_{1,0},\quad n=0,1,2,\dots
\end{equation}
provided $a_{2,0}^2+a_{1,0}^2\neq 0$ where the polynomials $y_n$, up to a multiplicative constant, may be readily obtained from the three-term recurrence relation:
\begin{align}\label{eq20}
&y_{n+2}=\bigg[A_nx+B_n\bigg]y_{n+1}+C_ny_n,\quad n\geq 0
\end{align} 
with the coefficients given by
\begin{align*}
A_n&={((2n+1)a_{2,0}+a_{1,0})(2(n+1)a_{2,0}+a_{1,0})\over (na_{2,0}+a_{1,0})},\\
B_n&={((2n+1)a_{2,0}+a_{1,0})(2n(n+1)a_{2,0}a_{2,1}+2(n+1)a_{1,0}a_{2,1}-2a_{1,1}a_{2,0}+a_{1,0}a_{1,1})\over (na_{2,0}+a_{1,0})(2na_{2,0}+a_{1,0})}\bigg],\\
C_n&={(n+1)(2(n+1)a_{2,0}+a_{1,0})((4a_{2,2}a_{2,0}^2-a_{2,0} a_{2,1}^2)n^2+
(4a_{2,0}a_{1,0}a_{2,2}-a_{1,0}a_{2,1}^2)n
+a_{1,0}^2a_{2,2}-a_{1,1}a_{1,0}a_{2,1}+a_{2,0}a_{1,1}^2)\over (na_{2,0}+a_{1,0})(2na_{2,0}+a_{1,0})},
\end{align*}
initiated with 
\begin{equation*}
y_0=1,\quad y_1=a_{1,0}x+a_{1,1}.
\end{equation*}

\noindent In the next sections, we shall apply the result of theorem~2 to study the possible quasi-exact analytic solutions for the $d$-dimension Schr\"odinger equation (\ref{eq3}) for unconstrained and constrained Coulomb plus harmonic oscillator potential (\ref{eq4}). We shall also apply AIM, theorem~1, to obtain \emph{accurate} approximations for arbitrary potential parameters, again, for the unconstrained and constrained $d$-dimensional Schr\"odinger equation (\ref{eq3}).
\section{Exact and approximate solutions for unconstrained potential $V(r)$}\label{spec}
\subsection{Exact bound-state solutions of a Coulomb plus 
harmonic oscillator potential in $d$-dimensions}

\noindent In this section, we consider the $d$-dimensional Schr\"odinger equation 
\begin{equation}\label{eq21}
\left[-{1\over 2}\left({d^2\over dr^2}-{(k-1)(k-3)\over 4r^2}\right)-{a\over r}+br^2\right]u_{nl}^d(r)=E_{nl}^d u_{nl}^d(r),\quad 0<r<\infty.
\end{equation}
In order to solve this equation by using AIM, the first step is to
transform (\ref{eq21}) into the standard form (\ref{eq5}). To this end, we note that the differential equation (\ref{eq21}) has one regular singular
point at $r = 0$ and an irregular singular point at $r = \infty$ and, since for large $r$, the harmonic oscillator term dominates, the asymptotic solution of (\ref{eq21}) as $r\rightarrow \infty$ is $u_{r\rightarrow \infty}\sim \exp({-{\sqrt{b/2}~r^2}})$; meanwhile the indicial equation of (\ref{eq21}) at the regular singular point $r=0$ yields
\begin{equation}\label{eq22}
s(s-1)-{1\over 4}(k-1)(k-3)=0,
\end{equation}
which is solved by
\begin{align*}
s_1={1\over 2}(3-k),\quad s_2={1\over 2}(k-1).
\end{align*}
The value of $s$, in Eq.(\ref{eq22}), determines the behavior of $u_{nl}^d(r)$ for $r\rightarrow 0$, and only $s>1/2$ is acceptable, since only in this case is the mean value of the kinetic energy finite \cite{landau}.
Thus, the exact solution of (\ref{eq21}) may assume the form
\begin{equation}\label{eq23}
u_{nl}^d(r)=r^{{1\over 2}(k-1)}\exp(-\sqrt{b\over 2}~r^2)~f_n(r),\quad k=d+3l,
\end{equation}
where we note that $u_{nl}^d(r)\sim r^{{1\over 2}(k-1)}$ as $r\rightarrow 0$. On substituting this ansatz wave function into (\ref{eq21}), we obtain the differential equation for $f_n(r)$ as
\begin{equation}\label{eq24}
rf_n''(r)+\left(-2 r^2\sqrt{2b}+k-1\right)f_n'(r)+\left[\left(2E_{nl}^d-k\sqrt{2b}\right)r+2a\right]f_n(r)=0.
\end{equation}
This equation is a special case of the differential equation (\ref{eq14}) with $a_{3,0}=a_{3,1}=a_{3,3}=a_{2,1}=0$, $a_{3,2}=1$, $a_{2,0}=-2 r^2\sqrt{2b},~a_{2,2}=k-1$, $\tau_{1,0}=-2E_{nl}^d+k\sqrt{2b}$ and $\tau_{1,1}=-2a$. Thus, the necessary condition for the polynomial solutions of Eq.(\ref{eq24}) is 
\begin{equation}\label{eq25}
E_{nl}^d=(2n'+k)\sqrt{b\over 2},\quad n'=0,1,2,\dots.
\end{equation}
and the sufficient condition follows from the vanishing of the tridiagonal determinant $\Delta_{n+1}=0$, $n=0,1,2,\dots$, namely 
\begin{center}
$\Delta_{n+1}$~~=~~\begin{tabular}{|lllllll|}
 $\beta_0~~$ & $\alpha_1$ &~&~& ~&~ &~\\
  $\gamma_1$ & $\beta_1$ &  $\alpha_2$&~&~&~&~ \\
~ & $\gamma_2$  & $\beta_2$&$\alpha_3$&~&~&~\\
$~$&~&$\ddots$&$\ddots$&$\ddots$&~&~\\ 
~&~&~&$\gamma_{n-2}$&$\beta_{n-2}$&$\alpha_{n-1}$&~\\
~&~&~&&~$\gamma_{n-1}$&$\beta_{n-1}$&$\alpha_n$\\
~&~&~&~&$~$&$\gamma_{n}$&$\beta_n$\\
\end{tabular}~~=~~0
\end{center}
where its entries are expressed in terms of the parameters of Eq.(\ref{eq24}) by
\begin{align}\label{eq26}
\beta_n&=-2a,\quad \alpha_n=-n(n+k-2),\quad \gamma_n=-2(n'-n+1)\sqrt{2b},
\end{align}
where $n'=n$ is fixed by the size of the determinant $\Delta_{n+1}=0$ and represent the degree of the polynomial solution of Eq.(\ref{eq24}). We may note that, since the off-diagonal entries $\alpha_i$ and $\gamma_i$ of the tridiagonal determinant satisfy the identity $$\alpha_i\gamma_i>0,\quad \forall~~ i=1,2,\dots,$$ the latent roots of the determinant $\Delta_{n+1}$ are all real and distinct \cite{arscott}.  Further, we can easily show that the determinant (\ref{eq26}) satisfies a three-term recurrence relation
\begin{equation}\label{eq27}
\Delta_i=\beta_{i-1}\Delta_{i-1}-\gamma_{i-1}\alpha_{i-1}\Delta_{i-2}, \quad \Delta_0=1,~~\Delta_{-1}=0,\quad i=1,2,\dots
\end{equation}
which can be used to compute the determinant $\Delta_i$ (and thus the sufficient conditions), recursively in terms of lower order determinants. In this case, however, we must fix $n'$ for each of the sub-determinants used in computing (\ref{eq27}). For example, in the case of $n'=n=1$ (corresponding to a polynomial solution of degree one),
we have 
\begin{center}
$\Delta_{2}$~~=~~\begin{tabular}{|ll|}
 $-2a$ & $-(k-1)$ \\
  $-2\sqrt{2b}$ & $-2a$ \\
\end{tabular}~~=~$\beta_{1}\Delta_{1}-\gamma_{1}\alpha_{1}\Delta_{0}$=$(-2a)(-2a)-(-2\sqrt{2b})(-(k-1))(1)$=$4a^2-2\sqrt{2b}(k-1),$
\end{center}
that is, the condition of the potential parameters reads
\begin{equation}\label{eq28}
2a^2-\sqrt{2b}(k-1)=0.
\end{equation}
For $n'=n=2$ (corresponding to a second-degree polynomial solution)
\begin{center}
$\Delta_{3}$~~=~~\begin{tabular}{|lll|}
 $-2a~~$ & $-(k-1)$ &$0$\\
  $-4\sqrt{2b}$ & $-2a$ &  $-2k$\\
0& $-2\sqrt{2b}$  & $-2a$
\end{tabular}~=$\beta_{2}\Delta_{2}-\gamma_{2}\alpha_{2}\Delta_{1}$=$(-2a)\begin{tabular}{|ll|}
 $-2a~~$ & $-(k-1)$ \\
  $-4\sqrt{2b}$ & $-2a$
\end{tabular}-(-2\sqrt{2b})(-2k)(-2a)$\\
=~$(-2a)[(-2a)\Delta_1-(-4\sqrt{2b})(-(k-1))\Delta_0$]+$8ka\sqrt{2b}$~=~$8a(-a^2+2\sqrt{2b}k-\sqrt{2b})$
\end{center}
Consequently, we must have
\begin{equation}\label{eq29}
a(a^2-2\sqrt{2b}k+\sqrt{2b})=0.
\end{equation}
In Table I, we give the conditions on the potential parameters to allow for polynomial solutions, from theorem 2. 

\vskip0.true in
\begin{table}[h] \caption{Conditions on the parameters $a$ and $b$ for the exact solutions of Eq.(\ref{eq21}) with $E_{nl}^d=(2n+k)\sqrt{b/2}$, $k=d+2l$. } 
\centering 
\begin{tabular}{l l l} 
\hline\hline 
$n$ & ~ &$\Delta_{n+1}=0 $\\ [0.5ex] 
\hline\hline 
0   &~& $a=0$\\ 
\hline 
1   &~& $2a^2 -(k -1)\sqrt{2b}=0$\\ 
\hline
2  &~& $a(a^2-(2k-1)\sqrt{2b})=0$\\ 
\hline
3   &~& $4a^4-20a^2\sqrt{2b}k-18b (1-k^2)=0$\\ 
\hline
4   & ~&$a(a^4-5\sqrt{2b}(2k+1)a^2+4b(8k^2+8k-7))=0$\\ 
\hline
5   & ~&$8a^6-140\sqrt{2b}(k+1)a^4+4b(-65+518k+259k^2)a^2-450b\sqrt{2b}(k-1)(k+3)(k+1)=0$\\ 
\hline
\hline 
\end{tabular} 
\label{table:nonlin}
\end{table}

\noindent It must be clear that although $n$, the degree of the polynomial solution, it is not necessarily an indication as to the number of the zeros of the wave function (node number): further analysis of the roots of $f_n(r)$ is usually needed to compute the zeros of the wavefunction. 
\vskip0.1true in
\noindent The polynomial solutions $f_{n'}(r)=\sum_{i=0}^{n'} c_ir^i$ can be easily constructed for each $n'$ since, in this case, the coefficients $c_i$ satisfy the three-term recurrence relation (see Eq.(\ref{eq17}))
\begin{equation}\label{eq30}
 c_{-1}=0,\quad c_0=1,\quad c_{i+1}=-{2ac_i+2(n'-i+1)\sqrt{2b}c_{i-1}\over (i+1)(i+k-1)},~~ i=0,1,\dots,n'-1,
\end{equation}
where $n'$ is the degree of the polynomial solution. When $n'=0$, $f_0(r)=1$.
For the first-degree polynomial solution, $n'=1,i=0$, we have
$$c_1=-{2a\over k-1},
$$
that is 
\begin{equation}\label{eq31}
f_1(r)=1-{2a\over k-1} r,\quad \mbox{where}\quad 2a^2 -(k -1)\sqrt{2b}=0.
\end{equation}
We may further note for $a<0$, there is no root of $f_1(r)=0$ and the un-normalized wave function reads
\begin{equation}\label{eq32}
u_{0l}^d(r)=r^{{1\over 2}(d+2l-1)}\exp\left(-{a^2r^2\over d+2l-1}\right)~\left(1-{2a r\over d+2l-1}\right),\quad a<0
\end{equation}
which represents a ground-state wave function in every subspace labeled by $d$ and $l$. For $a>0$, there is only one root of $f_1$ and the wave function
\begin{equation}\label{eq33}
u_{1l}^d(r)=r^{{1\over 2}(d+2l-1)}\exp\left(-{a^2r^2\over d+2l-1}\right)~\left(1-{2a r\over d+2l-1}\right),\quad a>0
\end{equation}
which represents a first excited-state in each subspace labeled by $d$ and $l$. The zero of this wave function is located at
\begin{equation}\label{eq34}
R={k-1\over 2a}, \quad k=d+2l,\quad a>0.
\end{equation}
In both cases, $a>0$ or $a<0$, the exact eigenvalues are given by
\begin{equation}\label{eq35}
E_{0l}^d\equiv E_{1l}^d\equiv E_{1l\pm 1}^{d\mp 2}={a^2(d+2l+2)\over (2d+4l-1)},\quad\quad \lim_{d\rightarrow \infty}E_{1l\pm 1}^{d\mp 2}={a^2\over 2}.
\end{equation}

\noindent For second-degree polynomial solution, $n'=2,i=0,1$, we have for the polynomial solution, $f_2(r)=c_0+c_1r+c_2r^2$, coefficients 
\begin{align*}
c_0=1,\quad c_1&=-{2a\over k-1}\quad \mbox{and}\quad c_2={2(a^2-\sqrt{2b}(k-1))\over k(k-1)}
\end{align*}
and the polynomial solution then reads
\begin{equation}\label{eq36}
f_2(r)=1-{2a r\over (k-1)}+{2a^2r^2\over (k-1)(2k-1)},
\end{equation}
where $a(a^2-(2k-1)\sqrt{2b})=0$ from which we may conclude that $a^2-\sqrt{2b}(k-1)>0$. Therefore, the wave function $f_2(r)$ has either two roots or no root based on the value of $a>0$ or $a<0$, respectively. 
For $a>0$, we have a second-excited state wave function
\begin{equation}\label{eq37}
u_{2l}^d(r)=r^{{1\over 2}(d+2l-1)}\exp\left(-{a^2r^2\over 2(2d+4l-1)}\right)~\left(1-{2a r\over d+2l-1}+{2a^2r^2\over (d+2l-1)(2d+4l-1)}\right),\quad a>0,
\end{equation}
which has two zeros at 
\begin{equation}\label{eq38}
R_1={2k-1+\sqrt{2k-1}\over 2a},\quad R_2={2k-1-\sqrt{2k-1}\over 2a},\quad k=d+2l,\quad a>0.
\end{equation}
For $a<0$, we have a ground-state wave function
\begin{equation}\label{eq39}
u_{2l}^d(r)=r^{{1\over 2}(d+2l-1)}\exp\left(-{a^2~r^2\over 2(2d+4l-1)}\right)~\left(1-{2a r\over d+2l-1}+{2a^2r^2\over (d+2l-1)(2d+4l-1)}\right),\quad a<0.
\end{equation}
In either case, $a>0$ or $a<0$, the exact eigenvalues reads
\begin{equation}\label{eq40}
E_{0l}^d\equiv E_{2l}^d\equiv E_{2l\pm 1}^{d\mp 2}={a^2(d+2l+4)\over 2(2d+4l-1)},\quad\quad \lim_{d\rightarrow \infty}E_{2l\pm 1}^{d\mp 2}={a^2\over 4}.
\end{equation}
For third-degree polynomial solution, $n'=3,i=0,1,2$, we have for the polynomial coefficients $f_3(r)=c_0+c_1r+c_2r^2+c_3r^3$ that
\begin{align*}
c_0=1,\quad c_1&=-{2a\over k-1},\quad c_2={(2a^2-3\sqrt{2b}(k-1))\over k(k-1)},\quad c_3=-{2a(2a^2-7\sqrt{2b}k+3\sqrt{2b})\over 3(k-1)k(k+1)},
\end{align*}
and the polynomial solution then reads
\begin{equation}\label{eq41}
f_3(r)=1-{2a\over k-1}r+{(2a^2-3(k-1)\sqrt{2b})\over k(k-1)}r^2-{2a(2a^2-(7k-3)\sqrt{2b})\over 3(k-1)k(k+1)}r^3,
\end{equation}
where the potential parameters satisfy the condition $4a^4-20a^2\sqrt{2b}k+18b (k^2-1)=0$ which may be solved in terms of $\sqrt{2b}$ as
\begin{equation}\label{eq42}
\sqrt{2b}={2a^2(5k\pm \sqrt{16k^2+9})\over 9(k^2-1)}.
\end{equation}
From this we have
\begin{equation}\label{eq43}
f_3^+(r)=1-{2ar\over k-1}-{2a^2\over 3}{(2k-3+\sqrt{16k^2+9})r^2\over (k+1)k(k-1)}+{4\over 9}{a^3(26k^2-15k+9+(7k-3)\sqrt{16k^2+9})r^3\over 3(k-1)^2k(k+1)^2},
\end{equation}
and
\begin{equation}\label{eq44}
f_3^-(r)=1-{2ar\over k-1}-{2a^2\over 3}{(2k-3-\sqrt{16k^2+9})r^2\over (k+1)k(k-1)}+{4\over 9}{a^3(26k^2-15k+9-(7k-3)\sqrt{16k^2+9})r^3\over 3(k-1)^2k(k+1)^2}.
\end{equation}
The polynomial $f_3^+(r)$ has two roots if $a>0$ and only one root if $a<0$ for all $r>0$; while $f_3^-(r)$ has no root for $a<0$ and has three roots for $a>0$ (the results that follow from Descartes' rule of signs). In each of these cases, the eigenvalues are given by
\begin{equation}\label{eq45}
E_{3l}^{d\pm}={a^2\over 9}{(d+2l+6)\left(5d+10l\pm \sqrt{16(d+2l)^2+9}\right)\over (d+2l-1)(d+2l+1)},\quad a\neq 0.
\end{equation}
We can also show for the fourth-degree polynomial solution, $n'=4,i=0,1,2,3$, we have
\begin{equation}\label{eq46}
f_4(r)=1-{2a\over k-1}r+{(2a^2-4\sqrt{2b}(k-1))\over k(k-1)}r^2+{4a[-a^2+5\sqrt{2b}k-2\sqrt{2b}]\over 3(k-1)k(k+1)}r^3+{2[a^4-(1+8k)\sqrt{2b}a^2+12b(k^2-1)]\over 3(k+2)(k+1)k(k-1)}r^4
\end{equation}
subject to $
a(a^4-5\sqrt{2b}(2k+1)a^2+4b(8k^2+8k-7))=0$ and in this case
\begin{equation}\label{eq47}
E_{4l}^{d\pm}={a^2\over 8}{(k+8)\left(10k+5\pm 3\sqrt{(2k+1)^2+8}\right)\over 2(2k+1)^2-9},\quad k=d+2l,a\neq 0
\end{equation}
and similarly for other cases. Indeed, using the recurrence relation, Eq.(\ref{eq30}), it is straightforward to compute explicitly the polynomial solution of any required degree.

\subsection{Approximate solutions for arbitrary potential parameters on half-line}
\noindent For arbitrary values of the potential parameters $a$ and $b$ that do not necessarily obey the above conditions, we may use AIM directly to compute the eigenvalues \emph{accurately},  as the zeros of the termination condition (\ref{eq13}). The method can be used, as well, to test the exact solutions we obtained in the above section. To utilize AIM, we start with
\begin{equation}\label{eq48}
\left\{ \begin{array}{l}
\lambda_0(r)=2 \sqrt{2b}~r-{(k-1)\over r}, \\ \\
  s_0(r)= - {(2E_{nl}^d-k\sqrt{2b})}-{2a\over r}
       \end{array} \right.
\end{equation} 
and computing the AIM sequences $\lambda_n$ and $s_n$  as given by Eq.(\ref{eq10}). We should note that for given values of the potential parameters $a$, $b,$ and of $k=d+2l$, the termination condition  $\delta_n=\lambda_n s_{n-1}-\lambda_{n-1}s_n=0
$ yields an expression that depends on both $r$ and $E$. In order to use AIM as an approximation technique for computing the eigenvalues $E$ we need to feed AIM with an initial value of $r=r_0$ that could stabilize AIM (that is, to avoid oscillations). For our calculations, we have found that $r_0=3$ stablizes AIM and allows us to compute the eigenvalues for arbitrary $k=d+2l$ and $n$ as shown in Table \ref{table:difdn}. There is no magical assertion about $r_0=3$, indeed using an exact solvable case, say $E=2.5$ with $d=3,l=0, n'=1$ for $a=1$ and $b=1/2$, we may approximate $r=r_0$ by means of $E-V(r)=0$ which yields $r_0\sim 2.4$ as an initial starting value for the AIM process. The eigenvalue computations in Table \ref{table:difdn}  were done using Maple version 13 running on an IBM architecture personal computer, where we used a high-precision environment. In order to accelerate our computation we have written our own code for a root-finding algorithm instead of using the default procedure {\tt Solve} of \emph{Maple 13}.

\begin{table}[h] \caption{Eigenvalues $E_{n0}^{d=2,3,4,5,6,7}$ for $V(r)=-1/r+r^2/2$. The initial value utilize AIM is $r_0=3$. The subscript $N$ refer to the number of iteration used by AIM.\\ } 
\centering 
\begin{tabular}{|c|p{2.5in}||c|p{2.5in}|}
\hline
$n$&$E_{n0}^{d=2}$&$n$&$E_{n0}^{d=3}$\\ \hline
0&$-1.836~207~439~051~476~488_{N=78}$&0&$~0.179~668~484~653~553~873_{N=71}$\\
1&$~~1.576~895~542~024~474~773_{N=71}$&1&$~2.500~000~000~000~000~000_{N=3~Exact}$\\
2&$~~3.828~388~290~161~145~035_{N=64}$&2&$~4.631~952~408~873~053~214_{N=59}$\\
3&$~~5.963~137~645~125~787~098_{N=60}$&3& $~6.712~595~725~661~429~760_{N=58}$ \\
4&$~~8.052~626~115~348~259~660_{N=59}$&4&$~8.769~519~600~328~899~714_{N=57}$\\
5&$~10.118~396~975~257~306~974_{N=58}$&5&$10.812~924~292~726~383~736_{N=56}$\\
6&$~12.169~728~962~611~565~630_{N=58}$&6&$12.847~666~480~105~796~414_{N=55}$\\ \hline
$n$&$E_{n0}^{d=4}\equiv E_{n1}^{2}$&$n$&$E_{n0}^{d=5}\equiv E_{n1}^{3}$\\ \hline
0&$~1.039~629~453~693~666~062_{N=63}$&0&$~1.709~018~091~123~552~219_{N=60}$\\
1&$~3.191~127~807~756~594~984_{N=58}$&1&$~3.801~929~609~626~278~046_{N=55}$\\
2&$~5.273~870~062~099~308~315_{N=57}$&2&$~5.860~357~172~819~176~603_{N=55}$\\
3&$~7.329~588~502~119~331~779_{N=54}$&3& $~7.902~317~748~608~790~676_{N=52}$ \\
4&$~9.371~002~235~676~830~145_{N=53}$&4&$~9.934~707~216~855~127~521_{N=51}$\\
5&$11.403~631~794~919~620~038_{N=53}$&5&$11.960~878~210~587~747~585_{N=51}$\\
6&$13.430~355~223~285~870~950_{N=53}$&6&$13.982~705~338~927~982~296_{N=50}$\\
\hline
$n$&$E_{n0}^{d=6}\equiv E_{n1}^{4}\equiv E_{n2}^{2}$&$n$&$E_{n0}^{d=7}\equiv E_{n1}^{5}\equiv E_{n2}^{3}$\\ \hline
0&$~2.311~633~609~259~797~633_{N=56}$&0&$~2.882~228~025~698~769~118_{N=52}$\\
1&$~4.376~287~059~247~773~643_{N=52}$&1&$~4.930~673~420~047~524~772_{N=50}$\\
2&$~6.420~575~455~465~976~803_{N=52}$&2&$~6.965~837~318~124~071~248_{N=48}$\\
3&$~8.453~864~208~404~376~310_{N=49}$&3& $~8.993~183~541~000~280~301_{N=47}$ \\
4&$~10.480~309~010~248~775~417_{N=50}$&4&$~11.015~405~735~685~306~483_{N=47}$\\
5&$12.502~107~717~572~917~269_{N=48}$&5&$13.034~024~645~001~717~876_{N=47}$\\
6&$14.520~559~118~487~876~168_{N=48}$&6&$15.049~980~236~362~344~263_{N=47}$\\
\hline

\end{tabular}
\label{table:difdn}
\end{table}

\section{Exact and approximate solutions for constrained potential}\label{spec}
\subsection{Analytic solutions}
\noindent We now consider the $d$-dimensional Schr\"odinger equation 
\begin{equation}\label{eq49}
\left[-{1\over 2}\left({d^2\over dr^2}-{(k-1)(k-3)\over 4r^2}\right)+V(r)\right]u_{nl}^d(r)=E_{nl}^d u_{nl}^d(r),\quad 0<r<R,
\end{equation}
where
\begin{equation}\label{eq50}
V(r)=\left\{ \begin{array}{ll}
 -{a\over r}+br^2, &\mbox{ if $0<r<R$} \\ \\
  \infty &\mbox{ if $r\geq R$}
       \end{array} \right.
\end{equation}
and $u_{nl}^d(0)=u_{nl}^d(R)=0$. We may assume the following ansatz for the wave function
\begin{equation}\label{eq51}
u_{nl}^d(r)=r^{{1\over 2}(k-1)}(R-r)\exp\left(-\sqrt{b\over 2}~r^2\right)~f_n(r),\quad k=d+2l.
\end{equation}
where $R$ is the radius of confinement, and the $(R-r)$ factor ensures that the radial wavefunction $u_{nl}^d(r)$ vanishes at the boundary $r=R$. On substituting (\ref{eq42}) into (\ref{eq40}), we obtain the following second-order differential equation for the functions $f_n(r)$:
\begin{align}\label{eq52}
f_n''(r)&=-2\left({k-1\over 2r}-{1\over R-r}-\sqrt{2b}~ r\right)f_n'(r)\notag\\
&-{1\over r(R-r)}\bigg[(-2E_{nl}^d+(k+2)\sqrt{2b})r^2+(R(2E_{nl}^d-k\sqrt{2b})-2a)r-k+1+2Ra\bigg]f_n(r).
\end{align}
We note that this equation reduces to Eq.(\ref{eq24}) as $R\rightarrow \infty$. Equation (\ref{eq52}) can be written as 
\begin{align}\label{eq53}
[-r^2+Rr]f_n''(r)&+[2\sqrt{2b}r^3-2\sqrt{2b}Rr^2-(k+1)r+(k-1)R]f_n'(r)\notag\\
&+{\bigg[(-2E_{nl}^d+(k+2)\sqrt{2b})r^2+(R(2E_{nl}^d-k\sqrt{2b})-2a)r+2Ra-k+1\bigg]}f_n(r)=0
\end{align}
This differential equation cannot be studied using Theorem 2. Consequently a further investigation of the following class of differential equations 
\begin{align}\label{eq54}
(a_{4,0}r^4+a_{4,1}r^3+a_{4,2}r^2&+a_{4,3}r+a_{4,4})y^{\prime \prime}+(a_{3,0}r^3+a_{3,1}r^2+a_{3,2}r+a_{3,3})y'-(\tau_{2,0}r^2+\tau_{2,1} r+\tau_{2,2})y=0,
\end{align}
is needed.  Indeed, by using Theorem 1 and a proof along the lines of the  proof of Theorem 2, we are able to establish the following:
\vskip0.1true in
\noindent{\bf Theorem 3.} \emph{
The second-order linear differential equation (\ref{eq54})
has a polynomial solution 
$y(r)=\sum_{k=0}^nc_k r^k$
if
\begin{equation}\label{eq55}
\tau_{2,0}=n(n-1)~a_{4,0}+n~a_{3,0},\quad n=0,1,2,\dots,
\end{equation}
provided $a_{4,0}^2+a_{3,0}^2\neq 0$ where the polynomial coefficients $c_n$ satisfy the five-term recurrence relation
\begin{align}\label{eq56}
((n-2)&(n-3)a_{4,0}+(n-2) a_{3,0}-\tau_{2,0})c_{n-2}+((n-1)(n-2)a_{4,1}+ (n-1)a_{3,1}-\tau_{2,1})c_{n-1}\notag\\
&+(n(n-1)a_{4,2}+ na_{3,2}-\tau_{2,2})c_n+(n(n+1)a_{4,3}+(n+1) a_{3,3} )c_{n+1}+(n+2)(n+1)a_{4,4}c_{n+2}=0
\end{align}
with
$c_{-2}=c_{-1}=0$. 
\vskip0.1true in
\noindent In particular, for the zero-degree polynomials $c_0\neq 0$ and $c_n=0,~n\geq 1$, we have 
\begin{equation}\label{eq57}
\tau_{2,2}=0,~\tau_{2,1}=0,~\tau_{2,0}=0.
\end{equation} 
For the first-degree polynomial solution,  $c_0\neq 0,\quad c_1\neq 0$ and $c_n=0,n\geq 2$, we must have 
\begin{equation}\label{eq58}
\tau_{2,0}=a_{3,0}
\end{equation}
along with the vanishing of the two $2\times 2$-determinants, simultaneously,
\begin{equation}\label{eq59}
\left|\begin{array}{ccc}
-\tau_{2,2}&a_{3,3}\\
-\tau_{2,1}& a_{3,2} -\tau_{2,2}
\end{array}\right|=0,\quad
\mbox{and}\quad
\left|\begin{array}{ccc}
-\tau_{2,2}&a_{3,3}\\
-a_{3,0}& a_{3,1} -\tau_{2,1}
\end{array}\right|=0.
\end{equation}
For the second-degree polynomial solution, $c_0\neq0,c_1\neq0,c_2\neq0$ and $c_n=0$ for $n\geq 3$, we must have
\begin{equation}\label{eq60}
\tau_{2,0}=2~a_{4,0}+2a_{3,0}
\end{equation}
along with the vanishing of the two $3\times 3$-determinants, simultaneously,
\begin{equation}\label{eq61}
\left|\begin{array}{ccc}
-\tau_{2,2}&a_{3,3}&2a_{4,4}\\
-\tau_{2,1}& a_{3,2} -\tau_{2,2}&2a_{4,3}+ 2a_{3,3}\\
-2a_{4,0}-2a_{3,0}&a_{3,1} -\tau_{2,1}&2a_{4,2}+2a_{3,2}-\tau_{2,2}
\end{array}\right|=0,~\mbox{and}~
\left|\begin{array}{ccc}
-\tau_{2,2}&a_{3,3}&2a_{4,4}\\
-\tau_{2,1}& a_{3,2} -\tau_{2,2}&2a_{4,3}+ 2a_{3,3}\\
0&-2a_{4,0}-a_{3,0}&2a_{4,1} +2a_{3,1}-\tau_{2,1}
\end{array}\right|=0,
\end{equation} 
and so on, for higher-order polynomial solutions. The vanishing of these determinants can be regarded as the conditions under which the coefficients $\tau_{2,1}$ and $\tau_{2,2}$ of Eq.(\ref{eq54}) are determined.
}
\vskip0.1true in
\noindent Using Theorem 3, we may note, with $a_{4,0}=a_{4,1}=a_{4,4}=0,a_{4,2}=-1,a_{4,3}=R,a_{3,0}=2\sqrt{2b},a_{3,1}=-2\sqrt{2b}R,a_{3,2}=-(k+1),a_{3,3}=(k-1)R,\tau_{2,0}=2E_{nl}^d-(k+2)\sqrt{2b},\tau_{2,1}=-(R(2E_{nl}^d-k\sqrt{2b})-2a),\tau_{2,2}=-2Ra+k-1$, that the necessary condition for polynomial solutions $f_n(r)=\sum_{k=0}^n c_k r^k$ of Eq.(\ref{eq53}) is 
\begin{equation}\label{eq62}
E_{nl}^d={1\over 2}(2n+k+2)\sqrt{2b},\quad k=d+2l,
\end{equation}
where $n$ refers to the degree of the polynomial solution and not necessarily to the number of zeros for the exact wave function. For sufficient conditions, we have for the zero-degree polynomial solution $n=0$, that Eq.(\ref{eq57}) yields 
\begin{equation}\label{eq63}
f_0(r)=1,\quad  E_{0l}^d={1\over 2}(k+2)\sqrt{2b},\quad \mbox{ if}\quad a=R\sqrt{2b}\quad\mbox{and}\quad Ra={1\over 2}(k-1), 
  \end{equation}
where, again, $k=d+2l$. For example, if $a=3,~b=4.5$, we have $R=1$ and for $k=d+2l=7$, we have the exact solution 
$$E_{00}^7=E_{01}^5=E_{02}^3=13.5$$
and for $a=4,b=8$, we have $R=1$ and for $k=d+2l=9$, we have the exact solution
$$E_{00}^9=E_{01}^7=E_{02}^5=E_{03}^3=22.$$
Thus,  for the values of the potential parameters $a,b$ and $R$ as given by 
\begin{equation}\label{eq64}
(a,b,R)=\left({1\over 2R}(2l+d-1),{1\over 8R^4}(2l+d-1)^2,R\right),
\end{equation}
we have the exact solutions
\begin{equation}\label{eq65}
\left\{ \begin{array}{l}
 E_{0l}^d={1\over 4R^2}(d+2l-1)(d+2l+2),\\ \\
  u_{0l}^d(r)=r^{{1\over 2}(2l+d-1)}(R-r)\exp(-{d+2l-1\over 4R^2}~r^2).
       \end{array} \right.
\end{equation}
We may note that the confinement size $R=(k-1)/(2a)$ represent the root of the unconfined wave function, (\ref{eq33}), with the same energy (compare (\ref{eq35} with (\ref{eq63})). 
\vskip0.1true in

\noindent For first-degree polynomial solution $n=1$, we have using (\ref{eq62}), or  $\tau_{2,0}=4\sqrt{2b}$,
\begin{equation}\label{eq66}
E_{1l}^d={1\over 2}(k+4)\sqrt{2b},
\end{equation}
along with the two conditions, obtained using (\ref{eq59}), which relate the potential parameters by
\begin{equation}\label{eq67}
\left\{ \begin{array}{l}
2kRa-k(k-1)-2R^2a^2+2R^2\sqrt{2b}(k-1)=0,\\ \\
2\sqrt{2b}R^2-2Ra+k-1=0.
       \end{array} \right.
\end{equation}
where, in this case, the polynomial solution reads
\begin{equation}\label{eq68}
f_1(r)=1-{(2Ra+1-k)\over R(k-1)}r.
\end{equation}
Thus, for the relations
\begin{equation}\label{eq69}
(a,b,R)=\left({2k-1+\sqrt{2k-1}\over 2R},{(k+\sqrt{2k-1})^2\over 8R^4},R\right),\quad k=d+2l
\end{equation}
we have the exact solutions
\begin{equation}\label{eq70}
\left\{ \begin{array}{l}
 E_{1l}^d={1\over 4R^2}(k+4)(k+\sqrt{2k-1}),\\ \\
  u_{1l}^d(r)=r^{{1\over 2}(k-1)}(R-r)\exp(-{k+\sqrt{2k-1}\over 4R^2}~r^2)\left(1-{(k+\sqrt{2k-1})\over R(k-1)}r\right).
       \end{array} \right.
\end{equation}
and for
 \begin{equation}\label{eq71}
(a,b,R)=\left({2k-1-\sqrt{2k-1}\over 2R},{(k-\sqrt{2k-1})^2\over 8R^4},R\right),\quad k=d+2l,
\end{equation}
we have the exact solutions
\begin{equation}\label{eq72}
\left\{ \begin{array}{l}
 E_{1l}^d={1\over 4R^2}(k+4)(k-\sqrt{2k-1}),\\ \\
  u_{1l}^d(r)=r^{{1\over 2}(k-1)}(R-r)\exp(-{k-\sqrt{2k-1}\over 4R^2}~r^2)\left(1-{(k-\sqrt{2k-1})\over R(k-1)}r\right).
       \end{array} \right.
\end{equation}
We note that these exact-solutions cases (\ref{eq69}) and (\ref{eq71}) represent the nodes of the wavefunction in the infinite case (\ref{eq37}).

\vskip0.1true in
\noindent For second-degree polynomial solutions $n=2$, we have the exact eigenvalues
\begin{equation}\label{eq73}
E_{nl}^d={1\over 2}(k+6)\sqrt{2b}
\end{equation}
where $k=d+2l$ and the potential parameters $a$, $b$ and $R$ are related by the following two conditions (obtained from the two determinants in (\ref{eq61})
\begin{equation}\label{eq74}
4R^3a^3-6(k+1)R^2a^2-2R(R^2\sqrt{2b}(7k-3)-3k(k+1))a+3(k-1)(k+1)(3\sqrt{2b}R^2-k)=0,
\end{equation}
and
\begin{equation}\label{eq75}
2R^2a^3-2R(R^2\sqrt{2b}+k)a^2-(k-1)(3\sqrt{2b}R^2-k)a+6b(k-1)R^3=0.
\end{equation}
In this case the exact solution reads
\begin{equation}\label{eq76}
u_{2l}^d=r^{{1\over 2} (k-1)} (R-r)\exp\left(-\sqrt{b\over 2} r^2\right) \left(1-{2Ra-k+1\over R(k-1)}r+{(2R^2a^2-2Rak+k(k-1)-3\sqrt{2b}R^2(k-1))\over R^2k(k-1)}r^2\right).
\end{equation}
Again in this case we can show that these exact solutions correspond to the zeros of the wavefunction in the infinite case (\ref{eq43}) and (\ref{eq44}).
\vskip0.1true in

\noindent Similar results can be obtained for higher $n$ (the degree of the polynomial solutions). It is important to note that the
conditions reported here are for the mixed potential $V (r) =-a/r + br^2$, where $a\neq 0$ and $b\neq  0$ (that is to say, neither coefficient is zero).
\subsection{Approximate solutions for confined potential with arbitrary parameters}
\noindent For the arbitrary values of $a, b$ and $R$, not necessarily satisfying the  above conditions, we may use AIM directly
to compute the eigenvalues with a very high degree of accuracy.  This also allows us to verify the exact solutions we obtained in the perevious sections. Similarly to the unconfined case, we start the iteration of the AIM sequence $\lambda_n$ and $s_n$ with
\begin{equation}\label{eq77}
\left\{ \begin{array}{l}
\lambda_0(r)=-2\left({k-1\over 2r}-{1\over R-r}-\sqrt{2b}~ r\right), \\ \\
  s_0(r)=-{(-2E_{nl}^d+(k+2)\sqrt{2b})r^2+(R(2E_{nl}^d-k\sqrt{2b})-2a)r+2Ra-k+1\over r(R-r)}.
       \end{array} \right.
\end{equation} 
where $0<r<R$. It is interesting to note in this case, that, unlike the unconfined case, the roots of the termination condition $\delta_n=\lambda_ns_{n-1}-\lambda_{n-1}s_n=0$ are much easier to handle in the present case. This is due to the fact that $r_0$ is now bound within $(0,R)$ for every given $R$. Thus, it is sufficient to start our iteration process with initial value $r_0=R/2$. In table \ref{table:difdn2},
we reported the eigenvalues we have computed using AIM for a fixed radius of confinement $R = 1$, with $r_0 = 0.5$ as an initial
value to seed the AIM process. In general, the computation of the eigenvalues is fast, as is illustrated by the small
number of iteration $N$ in Tables \ref{table:difdn2}. The same procedure can be applied to compute the eigenvalues for other values of $a$, $b$ $R$, and arbitrary dimension $d$. The results of AIM may be obtained to any degree of precision, although we have reported our results for only the first eighteen decimal places. It is clear from the table that our results confirm the invariance of the eigenvalues under the transformation $(d,l)\rightarrow (d\mp2,l\pm 1).$

\begin{table}[h] \caption{Eigenvalues $E_{nl}^{d=2,4}(a,b;R)$ for $V (r) = -a/r + br^2, r\in (0,R)$, where $a = \pm 1$, $b = 0.5$, $R = 1$ and different $n$ and $l$. The subscript $N$ refers to the number of iteration used by AIM.\\ } 
\centering 
\begin{tabular}{|c|c|p{2.2in}||c|c|p{2.2in}|}
\hline
$n$&$l$&$E_{nl}^{d=2}(1,1/2;1)$&$n$&$l$&$E_{nl}^{d=2}(1,1/2;1)$\\ \hline
0& 0&$-1.275~615~599~206~285~795_{N=32}$&0&0& $~-1.275~615~599~206~285~795_{N=32}$\\
&1&$~~5.400~467~192~272~980~536_{N=26}$&1&~&$~~10.924~630~155~130~440~587_{N=27}$\\
&2&$~11.652~661~600~597~110~050_{N=24}$&2&~&$~~32.734~045~433~763~800~052_{N=32}$\\
&3&$~19.010~259~174~813~201~428_{N=24}$&3&~&$~~64.522~506~980~951~712~401_{N=40}$\\
&4&$~27.565~689~679~299~850~255_{N=25}$&4&~&$~106.243~804~852~673~032~613_{N=47}$\\
&5&$~37.324~795~658~776~520~956_{N=28}$&5&~&$~157.875~359~994~443~580~341_{N=53}$\\
\hline\hline
$n$&$l$&$E_{nl}^{d=2}(-1,1/2;1)$&$n$&$l$&$E_{nl}^{d=2}(-1,1/2;1)$\\ \hline
0& 0&$~~6.107~045~323~129~696~121_{N=27}$&0&0&$~~6.107~045~323~129~696~121_{N=27}$\\
&1&$~~9.530~081~242~027~809~913_{N=24}$&1&~&$~19.534~700~629~074~427~546_{N=27}$\\
&2&$~15.106~527~319~660~138~719_{N=24}$&2&~&$~42.295~175~016~376~479~090_{N=35}$\\
&3&$~22.149~694~772~638~116~456_{N=24}$&3&~&$~74.728~314~736~027~030~722_{N=43}$\\
&4&$~30.518~339~762~183~359~381_{N=27}$&4&~&$~116.935~489~978~445~435~860_{N=47}$\\
&5&$~40.151~787~835~702~316~786_{N=29}$&5&~&$~168.956~853~183~684~793~313_{N=55}$\\
\hline\hline
$n$&$l$&$E_{nl}^{d=4}(1,1/2;1)$&$n$&$l$&$E_{nl}^{d=4}(1,1/2;1)$\\ \hline
0& 0&$~~5.400~467~192~272~980~536_{N=26}$&0&0&$~~~5.400~467~192~272~980~536_{N=24}$\\
&1&$~11.652~661~600~597~110~050_{N=24}$&1&~&$~~22.123~225~647~087~677~088_{N=25}$\\
&2&$~19.010~259~174~813~201~428_{N=24}$&2&~&$~~48.910~542~938~654~909~374_{N=35}$\\
&3&$~27.565~689~679~299~850~255_{N=25}$&3&~&$~~85.660~358~190~161~408~159_{N=43}$\\
&4&$~37.324~795~658~776~520~956_{N=28}$&4&~&$~132.333~925~295~766~891~686_{N=48}$\\
&5&$~48.278~874~241~139~597~779_{N=31}$&5&~&$~188.912~601~544~108~537~448_{N=54}$\\
\hline\hline
$n$&$l$&$E_{nl}^{d=4}(-1,1/2;1)$&$n$&$l$&$E_{nl}^{d=4}(-1,1/2;1)$\\ \hline
0& 0&$~~9.530~081~242~027~809~913_{N=24}$&0&0&$~~~9.530~081~242~027~809~913_{N=24}$\\
&1&$~15.106~527~319~660~138~719_{N=24}$&1&~&$~~27.374~386~080~371~192~265_{N=27}$\\
&2&$~22.149~694~772~638~116~456_{N=24}$&2&~&$~~54.884~084~396~689~521~442_{N=36}$\\
&3&$~30.518~339~762~183~359~381_{N=27}$&3&~&$~~92.165~200~694~649~737~766_{N=44}$\\
&4&$~40.151~787~835~702~316~786_{N=29}$&4&~&$~139.258~753~086~612~471~603_{N=49}$\\
&5&$~51.014~646~696~330~218~668_{N=32}$&5&~&$~196.184~615~317~703~801~052_{N=54}$\\
\hline
\end{tabular}
\label{table:difdn2}
\end{table}
\section{Conclusion}\label{conc}
We study a model atom-like system $-\half\Delta -a/r$ which is confined softly by the inclusion of a harmonic-oscillator potential term $b\,r^2$ and possibly also by the presence of an impenetrable spherical box of radius $R.$ For $b > 0$ or $R < \infty,$ the entire spectrum $E_{n,\ell}^d(a,b,R)$ is discrete.  We have studied these eigenvalues and we present an approximate spectral formula for the `free' case, $R = \infty$.  For the general case of $R\le\infty,$ AIM has been used to provide both a large number of exact analytical solutions, valid for certain special choices of the parameters $\{a,b,R\},$ and also very accurate numerical eigenvalues for arbitrary parametric data.
In the cases where we have found analytic solutions for $R=\infty$, the exact wave functions are no longer expressed in terms of known special functions, as is possible for the hydrogen atom.  However, the exact solutions we have found for confining potentials correspond to confinement at the zeros of the unconfined case. An interesting qualitative feature seems to be that $E_{n,\ell}^d(a,b,R)$, for large $R$, is concave with respect to $n$, $\ell$, 
or $d$, but becomes convex as $R$ is reduced; this may arise because the reduction in $R$ perturbs the higher states more severely since, when free, they are naturally  more spread out. It is hoped that the work reported in the present paper will provide a useful addition to the growing body of results concerning the spectra of confined atomic systems in $d$ dimensions.

\section{Acknowledgments}
\medskip
\noindent Partial financial support of this work under Grant Nos. GP3438 and GP249507 from the 
Natural Sciences and Engineering Research Council of Canada
 is gratefully acknowledged by two of us (RLH and NS). KDS thanks the Department of Science and Technology, 
New Delhi, for the J.C. Bose fellowship award. NS and KDS are grateful for the
  hospitality provided by the Department of Mathematics and Statistics of 
Concordia University, where part of this work was carried out.  

\end{document}